\documentclass[12pt]{article}

\usepackage[utf8]{inputenc}
\usepackage[T1]{fontenc}
\usepackage{lmodern}

\usepackage{xcolor}  
\usepackage[colorlinks=true,
            linkcolor=blue,
            citecolor=green!40!black,
            urlcolor=blue]{hyperref}

\usepackage[a4paper,margin=1in]{geometry}

\usepackage{amsmath, amssymb, amsthm, mathtools}

\newtheorem{definition}{\bf Definition}[section]

\newtheorem{result}{\bf Result}[section]
\newtheorem{corr}{\bf Corollary}[section]
\newtheorem{prop}{\bf Property}[section]
\newtheorem{rem}{\bf Remark}[section]

\usepackage{authblk}

\title{A Criterion for Safe Overshoot \\ in Coupled Tipping Systems}

\author[1,2]{S. Sinet\thanks{Corresponding author: s.a.m.sinet@uu.nl}}
\author[3]{N. A. M. Delmeire}
\author[4]{P. D. L. Ritchie}
\author[1,2]{H. A. Dijkstra}
\author[1,2]{A. S. von der Heydt}

\affil[1]{Department of Physics, Institute for Marine and Atmospheric Research Utrecht, Utrecht University, Utrecht, The Netherlands}
\affil[2]{Centre for Complex Systems Studies, Utrecht University, Utrecht, The Netherlands}
\affil[3]{Delft Institute of Applied Mathematics, Delft University of Technology, Delft, The Netherlands}
\affil[4]{Department of Mathematics and Statistics, Faculty of Environment, Science and Economy, University of Exeter, Exeter, UK}

\date{}

\begin{document}
\maketitle
\begin{abstract}
 Abrupt transitions are a central concern in climate and ecological research, and may arise when critical thresholds known as tipping points are crossed. However, previous work has shown that finite-time overshoots of tipping points can be safe, and that such behavior is captured by an inverse-square-law criterion when overshoots are sufficiently small and slow. So far studied in isolated systems with external drivers, (un)safe overshoots may also emerge from interactions between subsystems. Here, we investigate safe-overshoot phenomena in unidirectionally coupled slow-fast systems featuring both nonlinear interactions and coupling through time-derivatives. Specifically, we derive a criterion for the occurrence of safe overshoots analogous to the inverse-square law for isolated systems, but adapted to interactive settings, and expressed explicitly in terms of the timescale separation and coupling strength between subsystems. We illustrate these results using two conceptual models in which the Atlantic Meridional Overturning Circulation interacts with either the Amazon rainforest or the Greenland Ice Sheet.
\end{abstract}

\tableofcontents
\section{Introduction }\label{sec:Intro}
The study of tipping phenomena has gained increasing attention over the past decades across a wide range of domains. By tipping, we usually refer to an abrupt and often irreversible transition that occurs when external drivers push a system beyond a critical threshold, commonly called a tipping point \cite{lenton_tipping_2008}. This notion is particularly prominent in climate science, where the collapse of the Atlantic Meridional Overturning Circulation (AMOC) or large-scale disintegration of polar ice sheets are recognized as potential tipping points of the Earth system \cite{lenton_tipping_2008, armstrong_mckay_exceeding_2022}, and in ecology, where abrupt shifts including lake eutrophication, forest dieback, or mass coral-bleaching events have long been documented \cite{scheffer_catastrophic_2001}.

The common view of tipping is that it occurs when a dynamical bifurcation is crossed, a critical point beyond which the system undergoes qualitative change \cite{scheffer_catastrophic_2001,strogatz_nonlinear_2007}. The most studied case is fold (or saddle-node) bifurcations, beyond which the current stable state ceases to exist, resulting in a transition to another, often less desirable stable state. However, previous work has shown that overshooting such critical points during a finite time can be safe. Provided that they remain limited in magnitude and occur sufficiently slowly, the occurrence of such safe-overshoot phenomena is well captured by a simple criterion \cite{ritchie_inverse-square_2019}. More precisely, overshooting a fold bifurcation for a duration $D$ and with magnitude $M$ is safe whenever
\begin{equation}
	\sigma MD^2<16+\text{h.o.t.},
\end{equation}
where $\sigma$ is a system-dependent constant, and h.o.t. denotes higher-order terms that become negligible when the external driver evolves sufficiently slowly. The application of this criterion has been extensively examined in conceptual models of climate-system tipping elements, where safe overshoot may occur along global-warming trajectories that temporarily exceed the Paris Agreement target but return to it over longer timescales \cite{ritchie_overshooting_2021}.

While safe overshoot has been widely studied for isolated systems with external drivers, many real-world systems consist of multiple interacting components. In such settings, tipping-point overshoot may arise not only from external forcing, but also from interactions between subsystems. In this way, tipping in one subsystem may trigger transitions in others, leading to so-called cascading tipping events \cite{dekker_cascading_2018,klose_what_2021}. These cascades are of particular concern in the Earth system, where the AMOC, ice sheets, and other ecological components such as the Amazon rainforest evolve on widely different timescales and interact with varying strengths \cite{wunderling_climate_2024}, which may in turn amplify \cite{wunderling_interacting_2021, deutloff_high_2025} or, in some cases, dampen abrupt change \cite{sinet_amoc_2023,sinet_meltwater_2025, ciemer_impact_2021}. Although individual tipping points are generally considered low-likelihood events, and cascading tipping even more so, the possibility of their occurrence is alarming given the disproportionate consequences they may entail \cite{steffen_trajectories_2018}.

These considerations make it essential to investigate how safe overshoot operates in interacting tipping systems, where it may play a crucial role in reducing the likelihood of dangerous cascading events. Here, we demonstrate how safe overshoot can occur in interactive settings, and we provide a criterion for its occurrence analogous to the inverse-square law presented in \cite{ritchie_inverse-square_2019}.  In contrast to the single-system case, our criterion is formulated explicitly in terms of the timescale separation between the interacting subsystems, and the strength of their coupling. It applies to slow-fast systems with unidirectional interactions and, as in \cite{ritchie_inverse-square_2019}, to cases in which the overshoot is sufficiently slow and of sufficiently small magnitude.

The content of this paper is structured as follows. In Section \ref{sec:illustr}, we illustrate how safe and unsafe overshoots can arise in systems coupled via nonlinear terms and time derivatives. Section \ref{sec:Crit} presents a general criterion for distinguishing safe from unsafe overshoots in unidirectionally coupled slow-fast systems, along with a specific case relevant to the simplified models commonly used in (interacting) tipping-point research. These results are then illustrated in Section \ref{sec:Applications} using conceptual models of interacting tipping elements, including the AMOC-Amazon rainforest system and the AMOC-Greenland Ice Sheet system. Finally, Section \ref{sec:Conclusion} provides discussions and conclusions.

\section{Safe overshoot under interaction } \label{sec:illustr}
In this section, we illustrate how safe overshoot can occur in interactive settings, specifically in the case of coupled slow-fast systems with coupling only from the slow to the fast component. We will consider two examples that share the same slow state variable $x\in\mathbb{R}$, but differ in their fast state variables, either $y_1\in\mathbb{R}$ or $y_2\in\mathbb{R}$, and coupling functions, either $c_1:\mathbb{R}\rightarrow\mathbb{R}$ or $c_2:\mathbb{R}\rightarrow\mathbb{R}$. In the fast time $t$, the dynamics of these systems are given by 
\begin{align}
	\frac{1}{\epsilon}\dot{x} &= 3x-x^3+\mu,\label{eq:systexx}\\
	\dot{y}_{1,2} &= 2y_{1,2}^2 - 0.2 + \gamma c_\text{1,2}(x),\label{eq:systexy}
\end{align}
where
\begin{equation*}
c_1(x) = \frac{1}{5}\cos^2\left(\frac{\pi}{14}-\frac{2\pi x}{7}\right), \qquad c_2(x) = \dot{x}, 
\end{equation*}
and the dot denotes derivatives with respect to the fast time $t$. In both versions of this system, the slow variable $x$ evolves according to the same governing equation, where $\mu \in \mathbb{R}$ is a forcing parameter and $\epsilon>0$ sets the timescale separation between the slow and fast dynamics, such that system~\eqref{eq:systexx}-\eqref{eq:systexy} constitutes a slow-fast system when $\epsilon$ is small. The slow subsystem exhibits a double-fold bifurcation structure, with fold bifurcations occurring at the critical forcing parameter values $\mu_c = \pm 2$. The fast variables $y_1$ and $y_2$ share the same internal dynamics but differ in how they are coupled to the slow component: either through the nonlinear function $c_1(x)$ or through the time derivative $c_2(x)$. These choices are made to highlight important aspects that will become clear later. In both cases, the parameter $\gamma \in \mathbb{R}$ controls the strength of coupling, and each fast subsystem undergoes a fold bifurcation at the same critical value of the interacting term, namely when $\gamma c_1(x) = 0.2$ or $\gamma c_2(x) = 0.2$, respectively.

\begin{figure}[t!]
    \centering
    \includegraphics[scale=1.2]{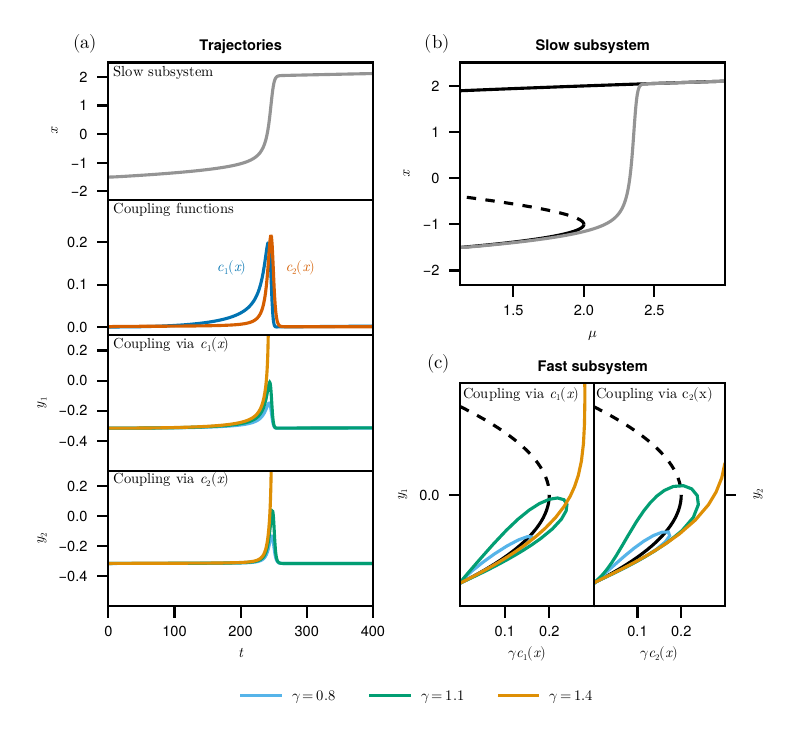}
    \caption{\emph{Trajectories and bifurcation structure of the two example systems defined by \eqref{eq:systexx}-\eqref{eq:systexy} under the slowly varying forcing \eqref{eq:muillustr}. Blue, green, and orange curves indicate trajectories for increasing coupling strengths $\gamma$, resulting in no overshoot, safe overshoot, and unsafe overshoot, respectively. \textnormal{(a)} From top to bottom: trajectory of the slow subsystem $x$; trajectory of the coupling functions $c_1(x)$ (dark blue) and $c_2(x)$ (dark orange); trajectories of the fast subsystem $y_1$ coupled via $c_1(x)$; and trajectories of the fast subsystem $y_2$ coupled via $c_2(x)$. \textnormal{(b)} Bifurcation diagram of the slow subsystem showing stable (solid black) and unstable (dashed black) equilibria. The slow-subsystem trajectory is overlaid in grey. \textnormal{(c)} Bifurcation diagrams of the fast subsystems $y_1$ (left) and $y_2$ (right), showing stable (solid black) and unstable (dashed black) equilibria. Fast-subsystem trajectories are overlaid in blue, green, and orange.}}
    \label{fig:Fig1}
\end{figure}

To investigate the transient response of these systems under a slow increase of the forcing, we replace $\mu$ by the (slow-)time-dependent function
\begin{equation}
    \mu(\epsilon t) = \mu_0 + r \epsilon t ,\label{eq:muillustr}
\end{equation}
where $\mu_0 = \tfrac{9}{8}$ and $r = 0.1$ determine the initial forcing level and its rate of increase in terms of the slow time $\epsilon t$, respectively, and we fix the timescale separation to $\epsilon = 0.05$. We initialise the system at $(x, y_{1,2}) = \left(-1.5, -\sqrt{0.1}\right)$, which is a stable equilibrium under the constant initial forcing $\mu = \mu_0$, and run the simulation until $t = 400$ for different values of the coupling strength $\gamma$. The resulting trajectories of the coupled systems are shown in Figure~\ref{fig:Fig1}a. As the slow subsystem crosses its fold bifurcation and tips (top panel; see also Figure~\ref{fig:Fig1}b), the corresponding trajectories of the coupling functions $c_1(x)$ and $c_2(x)$ are displayed in dark blue and dark orange, respectively (second panel from top). In both coupling scenarios, the final state of the fast subsystem depends on the value of the coupling strength $\gamma$ (Figure~\ref{fig:Fig1}a, lower panels, and Figure~\ref{fig:Fig1}c). In Figure~\ref{fig:Fig1}c, we find that for the smallest value of $\gamma$ (blue), no overshoot of the fast subsystem's fold bifurcation occurs. As $\gamma$ increases, we observe safe-overshoot trajectories in which the fast subsystem temporarily passes beyond the fold bifurcation, but eventually returns to its initial state (green). However, for the largest value of $\gamma$ (orange), the system fails to re-enter its original basin of attraction, resulting in tipping (i.e., unbounded growth in this case). 

To summarize, we find that the coupled system exhibits three distinct regimes depending on the coupling strength: no overshoot, safe overshoot, and unsafe overshoot of the fast subsystem's tipping point. We note, however, that the effect of varying the timescale separation has not yet been considered, even though it appears explicitly in the coupling function $c_2(x) = \dot{x} = \epsilon(3x - x^3 + \mu(\epsilon t))$. In the next section, we develop a criterion that determines whether a finite-time overshoot in such systems is safe or unsafe, and that clarifies how this outcome depends, in particular, on the coupling strength and the timescale separation.

\section{Criterion for safe overshoot}\label{sec:Crit}

In this section, we quantify how the interplay between coupling strength and timescale separation determines whether trajectories of the coupled system result in safe overshoot. First, in Section \ref{sec:CritGen}, we formulate a general criterion for the occurrence of safe overshoot in unidirectionally coupled slow-fast systems, applying to a large set of nonlinear interactions that also includes time derivatives. Second, in Section \ref{sec:CritSpec}, we formulate a corollary applying to simpler systems that are especially relevant to the conceptual study of interacting tipping elements. Finally, in Section \ref{sec:Appexamples}, we apply our theoretical results to the example systems used in Section \ref{sec:illustr}.

\subsection{General case}\label{sec:CritGen}
We consider a system of two state vectors, $\mathbf{x}\in\mathbb{R}^n$ and $\mathbf{y}\in\mathbb{R}^m$ with $n,m\in\mathbb{N}$, whose evolution with respect to the fast time $t$ is governed by
\begin{align} 
    \frac{1}{\epsilon}\dot{\mathbf{x}} &= f(\mathbf{x},\epsilon t), \label{eq:systgenx}\\
    \dot{\mathbf{y}} &= g\!\left(\mathbf{y},\, \gamma \epsilon^i c(\mathbf{x},\epsilon t)\right), \label{eq:systgeny}
\end{align}
where $f$ and $g$ are sufficiently smooth functions defining the dynamics of $\mathbf{x}$ and $\mathbf{y}$, respectively, and the dot denotes differentiation with respect to the fast time $t$. The parameter $\epsilon>0$ controls the timescale separation between the two subsystems, which exhibits slow-fast dynamics with slow component $\mathbf{x}$ and fast component $\mathbf{y}$ when $\epsilon$ is small. In \eqref{eq:systgeny}, the interaction is characterized by a sufficiently differentiable scalar coupling function $c$, a coupling parameter $\gamma\in\mathbb{R}\setminus\{0\}$ determining the strength and sign of the interaction, and a scaling $\epsilon^i$ with exponent $i\in\{0,1\}$. This formulation generalises the example systems considered in Section~\ref{sec:illustr}. The $\epsilon^i$ scaling allows us to include both nonlinear couplings that depend directly on $\mathbf{x}$ (i.e., in the case where $i=0$) and, in particular, interactions mediated by time derivatives of $\mathbf{x}$, which naturally scale as $\epsilon$ (i.e., in the case where $i=1$). Finally, we explicitly allow for non-autonomous effects by admitting a dependence on the slow time $\epsilon t$ in \eqref{eq:systgenx} and in the coupling function $c$.

We focus on cases where the fast subsystem undergoes a fold bifurcation at a critical value of the interaction component, and where the slow dynamics maximize this component at a finite time. Specifically, we make the following assumptions (throughout the text, a dot always denotes differentiation with respect to the fast time $t$, and $\partial_j\phi$ denotes the derivative of a function $\phi$ with respect to its $j^{\text{th}}$ argument):

\begin{description}
    \item[A1 (fold bifurcation).] At the critical value $b$ of the interaction component $\gamma \epsilon^{i} c(\mathbf{x},\epsilon t)$, the fast subsystem undergoes a fold bifurcation at $\mathbf{y} = \mathbf{y}_b$: 
\begin{itemize}
    \item 
    $g(\mathbf{y}_b,b)=0$, and the Jacobian matrix $\partial_1 g(\mathbf{y}_b,b)$ is singular with a one-dimensional kernel, 
    with a right null vector $\mathbf{v}_0$ and a left null vector $\mathbf{w}_0$,
    normalized so that $\mathbf{w}_0^\top \mathbf{v}_0 = 1$;
    
    \item 
    all other eigenvalues of $\partial_{1} g(\mathbf{y}_b,b)$ have strictly negative real parts;
    
    \item 
    $\alpha_0 := \mathbf{w}_0^\top \partial_2 g(\mathbf{y}_b,b) > 0$, 
    so that variation of the interaction component crosses the fold transversally;

    \item 
    $\kappa := \frac{1}{2\alpha_0}\,
    \mathbf{w}_0^\top \partial_1^{2} g(\mathbf{y}_b,b)\mathbf{v}_0^2 > 0$,
    ensuring that exactly one saddle and one node collide in the bifurcation.

\end{itemize}
    \item[A2 (Regular maximum of the interaction component).] Along the slow subsystem's trajectory $\mathbf{x}(t)$, the interacting component $\gamma \epsilon^i c(\mathbf{x}(t),\epsilon t)$ has a unique regular maximum at $t = t_{\max}$. We denote the value of $c$ at this maximum by $c_{\max}\equiv c\!\left(\mathbf{x}(t_{\max}),\epsilon t_{\max}\right)$. 
\end{description}

Note that, in Assumption~A2, while $t_{\max}$ depends on $\epsilon$, the product $\epsilon t_{\max}$ remains independent of $\epsilon$. This follows from the fact that the slow trajectory $\mathbf{x}(t)$ depends on time only through the slow-time variable $\epsilon t$. Consequently, $c(\mathbf{x}(t), \epsilon t)$ is a function solely of $\epsilon t$, such that the location of its maximum in slow time does not depend on $\epsilon$. Also, in the general definition of a fold bifurcation \cite{kuznetsov_elements_2004}, the sign of $\alpha_0$ and $\kappa$ is not prescribed. In Assumption A1, these signs have been fixed positive for convenience, in particular such that branches of equilibria connected to the fold exist for $\gamma \epsilon^{i} c(\mathbf{x},\epsilon t)<b$. Consistently, an overshoot of the fast subsystem's fold threshold $b$ can occur if $\gamma \epsilon^{i} c(\mathbf{x}(t),\epsilon t)$ attains a maximum along the slow trajectory, as stated in Assumption~A2. We formalize the concepts of overshoot, finite-time overshoot, and safe overshoot as follows:
\begin{definition}
\label{def:overshoot}
Let $b$ denote the fast subsystem's fold threshold from Assumption~A1. Along a trajectory $\mathbf{x}(t)$ of the slow subsystem satisfying Assumption~A2, we say that an overshoot (of the fold threshold $b$) occurs whenever
\begin{equation}
    \gamma \epsilon^{i} c_{\max}-b>0.
\end{equation}
We say that the overshoot is a finite-time overshoot if there exist $t_1<t_2$ such that
\begin{equation}
    \gamma \epsilon^{i} c\bigl(\mathbf{x}(t),\epsilon t\bigr)-b>0
    \quad \text{if and only if} \quad
    t\in(t_1,t_2).
\end{equation}
A finite-time overshoot is called safe if the fast subsystem trajectory $\mathbf{y}(t)$ lies in the basin of attraction of the attracting branch involved in the fold for at least some nontrivial finite time interval after $t_2$. Otherwise, the overshoot is called unsafe.
\end{definition}

Other sign conventions for $\alpha_0$ and $\kappa$, as well as cases in which an overshoot occurs at a minimum rather than a maximum of the interaction component, can be reduced to this setting by simple transformations, such as reversing the sign of $\mathbf{y}$, the coupling strength $\gamma$, or the coupling function $c$. 

As we have seen in Section~\ref{sec:illustr}, such scenarios can result in no overshoot, safe overshoot, or unsafe overshoot of the fast subsystem's fold bifurcation. We find that, for sufficiently strong timescale separation, and in cases where the overshoot occurs in the vicinity of the fast subsystem's fold bifurcation, a criterion for the occurrence of safe overshoots is given by the following result:

\begin{result}\label{res:1}  
Let the fast subsystem \eqref{eq:systgeny} satisfy A1, and let $\mathbf{x}(t)$ be a trajectory of the slow subsystem \eqref{eq:systgenx} satisfying A2. For sufficiently small $\epsilon$, an overshoot occurring near the fast subsystem's fold bifurcation (in the sense of Remark~\ref{rem:vicinity}) is safe whenever
\begin{equation}
    F \sqrt{\frac{-1}{\gamma \epsilon^{i+2} S}} \,\bigl(\gamma \epsilon^i c_{\max} - b\bigr) < 1 + o(1),\label{eq:safecritgen}
\end{equation}
where $F = \alpha_0 \sqrt{2\kappa}$, $o(1)$ denotes a quantity that vanishes as $\epsilon \to 0$, and 
\begin{equation}
    S 
    = \Bigl[
f^\top \bigl(\partial_{1}^2c\bigr)\, f
\;+\;
(\partial_{1}c)^\top (\partial_{1}f)\, f
\;+\;
(\partial_{1}c)^\top \partial_{2}f
\;+\;
2\,(\partial_{12}c)^\top f
\;+\;
\partial_{2}^2c
\Bigr]\left(\mathbf{x}(t_{\max}),\epsilon t_{\max}\right),\label{eq:S}
\end{equation}
is independent of $\epsilon$ and satisfies $\gamma S<0$.
\end{result}

We now provide a justification of Result~\ref{res:1}. The key steps are, first, to describe the interaction component in \eqref{eq:systgeny} in the vicinity of its maximum, and, second, to analyse the fast dynamics near the fold bifurcation. From there, the remainder of the argument follows closely the approach in \cite{ritchie_inverse-square_2019}. Our aim is to give an asymptotically consistent derivation rather than a fully rigorous proof. In particular, we build upon the assumption that solutions can be expressed as sufficiently regular expansions in the timescale separation $\epsilon$, and check consistency at leading order as $\epsilon\to 0$.

By Assumption A2, the interaction component has a unique regular maximum along the slow trajectory $\mathbf{x}(t)$. Without loss of generality, we shift time so that this maximum occurs at $t=0$ (i.e., $t_{\max} = 0$ in Assumption A2). Along this trajectory, noting that $\mathbf{x}(t)$ depends on $t$ only through the slow time $\epsilon t$, we rewrite the interaction component around its maximum as
\begin{equation}
    \gamma\epsilon^ic(\mathbf{x}(t),\epsilon t) = \gamma\epsilon^ic_{\max} + \gamma\epsilon^ih(\epsilon t), \label{eq:intcompqh}
\end{equation}
where the slow-time dependence is absorbed into a sufficiently smooth function $h$ which, by Assumption A2, satisfies $h(0)=\partial_1{h}(0)=0$ and $\gamma\partial_1^2{h}(0)<0$. 

Assuming that an overshoot of the fast subsystem's fold threshold $b$ occurs, we rewrite expression \eqref{eq:intcompqh} as 
\begin{equation}
    \gamma \epsilon^i c(\mathbf{x}(t),\epsilon t) = b + \left(\gamma \epsilon^i c_{\max} - b\right) + \gamma\epsilon^ih(\epsilon t),\label{eq:intcompbeforeR0}
\end{equation}
which separates three contributions: the position of the fold bifurcation $b$, the overshoot magnitude $\gamma \epsilon^i c_{\max} - b$, and the variation away from the maximum given by $\gamma\epsilon^ih(\epsilon t)$.

As in \cite{ritchie_inverse-square_2019}, our derivation of a safety criterion will rely on a local description of the fast dynamics near the fold bifurcation as $\epsilon\to 0$. 
To set up this local analysis, we introduce the following scalings, analogous to the approach of \cite{ritchie_inverse-square_2019}. On the one hand, we focus on the regime in which the overshoot magnitude is asymptotically small, and write
\begin{equation}
    \gamma \epsilon^{i} c_{\max} - b = R_0 \epsilon^{a_R}, \label{eq:scaleint}
\end{equation}
where $R_0>0$ is $\epsilon$-independent, and $a_R>0$ is an exponent that will be determined later.
This parametrization ensures that the overshoot magnitude decreases as the timescale separation increases (i.e., as $\epsilon$ decreases), a necessary condition for safe overshoot to occur. In particular, for sufficiently small $\epsilon$, it ensures that the overshoot will be finite in time. Substituting \eqref{eq:scaleint} into equation \eqref{eq:intcompbeforeR0} gives
\begin{equation}
    \gamma \epsilon^i c(\mathbf{x}(t),\epsilon t) = b + R_0 \epsilon^{a_R} + \gamma\epsilon^i h(\epsilon t), 
    \label{eq:intcompfinal}
\end{equation}
which is the form of the interaction component that we will use to analyse the fast dynamics near the fold bifurcation. On the other hand, we introduce the rescalings
\begin{align}
    z&=\epsilon^{-a_y}\mathbf{w}_0^\top(\mathbf{y} -\mathbf{y}_b), \label{eq:scalez} \\
    T &= \epsilon^{a_t} t, \label{eq:scaletau}
\end{align}
where $a_y$ and $a_t$ are positive exponents to be determined later (note that $T$ is a rescaled time that should not be confused with the slow time $\epsilon t$). For $\epsilon<1$, this change of variables zooms into a neighbourhood of the fold point $\mathbf{y}=\mathbf{y}_b$ along the critical direction, and accelerates time. Combining \eqref{eq:intcompfinal}-\eqref{eq:scaletau} with the fast dynamical equation \eqref{eq:systgeny} gives the scalar equation
\begin{equation}
    \epsilon^{a_y + a_t} z'
    = \mathbf{w}_0^\top g\!\left(
        \mathbf{y},\,
        b + R_0 \epsilon^{a_R} + \gamma \epsilon^i h\!\left(\epsilon^{1-a_t} T\right)
      \right),
    \label{eq:fastzoom_vec}
\end{equation}
where the prime denotes differentiation with respect to the rescaled time $T$. We now consider any bounded interval of the rescaled time (i.e.\ $T=O(1)$) and assume $a_t<1$ (which will be verified later). For sufficiently small $\epsilon$, we may use the Taylor expansion of $h$ around $0$. Since $h(0)=\partial_1 h(0)=0$, this yields
\begin{equation}
    h\!\left(\epsilon^{1-a_t}T\right)
    = \frac{1}{2}\partial_1^2h(0)\,\epsilon^{2(1-a_t)}T^2
      + O\!\left(\epsilon^{3(1-a_t)}\right),
    \label{eq:hexp}
\end{equation} 
where $\partial_1^2h(0)$ is $\epsilon$-independent. In what follows, we focus on the regime in which, as the interaction component $\gamma \epsilon^i c(\mathbf{x}(t),\epsilon t)$ approaches the critical value $b$, the fast variable remains close to the attracting equilibrium branch connected to the fold, so that the overshoot is initiated near the fold bifurcation. In particular, we assume that $z$ remains $O(1)$ (i.e.\ $\mathbf{w}_0^\top(\mathbf{y}(t)-\mathbf{y}_b)=O(\epsilon^{a_y})$). Since, by Assumption~A1, all modes that transverse the fold are exponentially stable, they relax rapidly compared to the slow variation of the interaction component. Thus, for sufficiently small $\epsilon$, contributions related to these modes only enter \eqref{eq:fastzoom_vec} at higher order while the trajectory remains in the fold neighbourhood. Consequently, a Taylor expansion of \eqref{eq:fastzoom_vec} about $(\mathbf{y}_b,b)$ together with \eqref{eq:hexp} yields the leading-order dynamics for $z$
\begin{equation}
    \epsilon^{a_y+a_t} z'
      = \alpha_0 \left[R_0 \epsilon^{a_R}
      + \frac{\gamma}{2}\partial_1^2h(0)\,\epsilon^{i+2(1-a_t)}T^2 + \kappa z^2\epsilon^{2a_y}\right]
        + O\left(\epsilon^E\right),
    \label{eq:normalform}
\end{equation}
where $\alpha_0$ and $\kappa$ are defined in Assumption~A1, and the order $E$ of the remainder is obtained by tracking all higher order corrections:
\begin{equation}
	E = \min\left(3a_y, 2a_R, a_y+a_R, a_y+2(1-a_t)+i, i+3(1-a_t)\right).
\end{equation}

At this stage, the dynamics still involve the three unknown exponents $a_R, a_y$ and $a_t$ introduced earlier. To fix these, we require that, as $\epsilon\rightarrow 0$, equation \eqref{eq:normalform} has a dominant balance at leading order, meaning that all the retained terms scale at the same order in $\epsilon$. The left-hand side scales like $\epsilon^{a_y+a_t}$. On the right-hand side, the three leading terms scale like $\epsilon^{a_R}$, $\epsilon^{i+2(1-a_t)}$, and $\epsilon^{2a_y}$. Hence, dominant balance yields the unique solution
\begin{equation}
    a_y = a_t = \tfrac{i+2}{4} \qquad\text{and}\qquad a_R = \tfrac{i+2}{2}.
\end{equation}
With this choice, all retained terms scale at order $(i+2)/2$ in $\epsilon$. We verify that, for both $i=0$ and $i=1$, the remainder term scales at strictly higher order (i.e., $E>(i+2)/2$), and that $a_t<1$, as required for the expansion \eqref{eq:hexp} of $h$ to apply. Substituting these values into equation \eqref{eq:normalform}, and dividing by $ \epsilon^{(i+2)/2}$, yields
\begin{align}
    z' &= \alpha_0\left(R_0 + \frac{\gamma}{2}\partial_1^2h(0) T^2 + \kappa z^2\right) + o(1),\label{eq:scalardiffeq}
\end{align}
where $o(1)$ denotes terms that vanish in the limit $\epsilon\rightarrow 0$. This problem is identical to the one considered in \cite{ritchie_inverse-square_2019}. In particular, for sufficiently small $\epsilon$, unsafe overshoot (i.e.\ tipping) corresponds to $z(T)$ growing to large positive values at a finite rescaled time $T$, which occurs when $R_0$ (encoding the overshoot magnitude) is sufficiently large. This is precisely the object of the inverse-square-law in \cite{ritchie_inverse-square_2019}, stating that, for sufficiently small $\epsilon$, the overshoot is safe whenever
\begin{equation}
    R_0 < \frac{1}{\alpha_0}\sqrt{-\frac{\gamma \partial_1^2h(0)}{2\kappa}} + o(1),
\end{equation}
which, in terms of the original variables and parameters, and using $\ddot{c}(\mathbf{x}(0),0) = \partial_1^2h(0)\epsilon^2$, yields 
\begin{equation}
    \left(\gamma \epsilon^i c_{\max}-b\right)\epsilon^{-\frac{i+2}{2}} < \frac{1}{\alpha_0}\sqrt{-\frac{\gamma\ddot{c}(\mathbf{x}(0),0)}{2\epsilon^2\kappa}} + o(1).
\end{equation}
Shifting back to the original time origin, computing the second time derivative gives  
\begin{equation}
    \ddot{c}(\mathbf{x}(t_{\max}),\epsilon t_{\max}) = \epsilon^2 S,\label{eq:Sdef}
\end{equation}
where
\begin{equation}
    S 
    = \Bigl[
f^\top \bigl(\partial_{1}^2c\bigr)\, f
\;+\;
(\partial_{1}c)^\top (\partial_{1}f)\, f
\;+\;
(\partial_{1}c)^\top \partial_{2}f
\;+\;
2\,(\partial_{12}c)^\top f
\;+\;
\partial_{2}^2c
\Bigr]\left(\mathbf{x}(t_{\max}),\epsilon t_{\max}\right).
\end{equation}
Substituting back into the criterion, and introducing the shorthand $F = \alpha_0 \sqrt{2\kappa}$, we recover
\begin{equation}
    F \sqrt{\frac{-1}{\gamma \epsilon^{i+2} S}} \,\bigl(\gamma \epsilon^i c_{\max} - b\bigr) < 1 + o(1),\label{eq:safecritgenjusti}
\end{equation}
which is criterion \eqref{eq:safecritgen}. To conclude the justification, we note that $S$ is independent of $\epsilon$, since both $\mathbf{x}(t_{\max})$ and $\epsilon t_{\max}$ are independent of $\epsilon$, and that the square root appearing in \eqref{eq:safecritgenjusti} is real, as $\gamma S=\gamma \epsilon^{-2}\ddot c(\mathbf{x}(t_{\max}),\epsilon t_{\max})<0$.

\begin{rem}\label{rem:vicinity}
In Result~\ref{res:1}, an overshoot is said to occur ``near the
fast subsystem's fold bifurcation'' if the scaling regime ensuring the validity of the local normal-form reduction \eqref{eq:normalform} as introduced above holds:
\begin{enumerate}
	\item as the interacting component
$\gamma \epsilon^i c(\mathbf{x}(t),\epsilon t)$ approaches the critical value $b$, the
fast variable remains sufficiently close to the attracting equilibrium branch connected to the fold with, in particular,
\begin{equation}
    \mathbf{w}_0^{\top}\left(\mathbf{y}(t) - \mathbf{y}_b\right) = O\!\left(\epsilon^{\frac{i+2}{4}}\right),
\end{equation}
	\item the overshoot magnitude is sufficiently small, satisfying
\begin{equation}
    \gamma \epsilon^i c_{\max} - b
      = O\!\left(\epsilon^{\frac{i+2}{2}}\right).
\end{equation}
\end{enumerate}
\end{rem}

Hence, this result provides a criterion for the occurrence of safe overshoot analogous to the formulation of \cite{ritchie_inverse-square_2019}, but adapted to interacting systems with, in particular, explicit dependence on the timescale separation $\epsilon$ and the interaction strength $\gamma$. Both the coefficients $F$ and $S$ appearing in criterion~\eqref{eq:safecritgen} are directly given when the governing equations are known, but can also be evaluated numerically. The coefficient $F$ can be obtained by numerical continuation via the following limit \cite{ritchie_inverse-square_2019}:
\begin{equation}
	F
	=
	\sqrt{
	\frac{1}{2}
	\lim_{\nu \to b}
	\frac{\lambda^2(\nu)}{\,b-\nu\,}
	},
	\label{eq:F}
\end{equation}
where $\nu$ denotes the interaction component in the fast subsystem
\eqref{eq:systgeny}, and $\lambda(\nu)$ is the leading (i.e., least negative)
eigenvalue of $\partial_1 g(\mathbf{y},\nu)$ evaluated along the fast subsystem's stable branch. Finally, given a numerical estimation of the second time derivative of $c$ evaluated at $t_{\max}$, $S$ can be evaluated via equation \eqref{eq:Sdef}.

\subsection{Simplified case: cubic slow system}\label{sec:CritSpec}
Having established the general criterion, we focus on a simplified case that is especially relevant to the conceptual study of interacting tipping elements. Namely, we assume that the slow component dynamics are autonomous and governed by a scalar cubic polynomial, which is widely used as a minimal representation of tipping dynamics \cite{wunderling_interacting_2021}. Second, we assume that the coupling from the slow to the fast component occurs through a time derivative. This is especially natural in the context of the interacting polar ice sheets and AMOC, where the slow dynamical variable often represents the ice sheet's mass or volume, and the interaction arises via meltwater fluxes that scale with their time derivative \cite{sinet_amoc_2023,sinet_amoc_2024,klose_rate-induced_2024}. The resulting system takes the form
\begin{align}
\frac{1}{\epsilon}\dot{x} &= a_3x^3+a_2x^2+a_1x+a_0, \label{eq:systP31}\\
\dot{\mathbf{y}} &= g(\mathbf{y},\gamma \dot{x}), \label{eq:systP32}
\end{align}
where $x\in\mathbb{R}$, $\mathbf{y}\in\mathbb{R}^m$ with $m\in\mathbb{N}$, $\epsilon>0$, and $a_3,a_2,a_1,a_0 \in \mathbb{R}$ with $a_3\neq 0$. We are interested in the case where the third degree polynomial has a double real root, as well as one simple real root corresponding to a stable equilibrium of equation \eqref{eq:systP31}. More precisely, we make the following assumption
\begin{description}
	\item[A3 (polynomial degeneracy).] Given equation \eqref{eq:systP31}, let
	\begin{align}
	p &= \frac{3a_3a_1-a_2^2}{3a_3^2},\label{eq:defp}\\ 
	q &= \frac{2a_2^3-9a_3a_2a_1+27a_3^2a_0} {27a_3^3}.\label{eq:defq}
	\end{align}
We assume that:
	\begin{itemize}
		\item $4p^3+27q^2 = 0$: the polynomial in \eqref{eq:systP31} has at least one repeated root,
		\item $p\neq 0$: the polynomial in \eqref{eq:systP31} has exactly two coinciding real roots, denoted $x_d$,  
		\item $a_3<0$: under the above conditions, the distinct simple real root denoted $x_s$ of the polynomial in \eqref{eq:systP31} corresponds to a stable equilibrium of equation \eqref{eq:systP31}.  
	\end{itemize}
\end{description}
This assumption places the slow subsystem in a configuration analogous to that of system \eqref{eq:systexx}-\eqref{eq:systexy} when the forcing parameter is fixed at a critical value corresponding to a fold bifurcation of the slow subsystem, where two equilibria coincide (at $x = x_d$) while an alternative stable state remains (at $x = x_s$). Under Assumption A3, the slow-fast system can be rewritten in a canonical form by an appropriate change of variable and parameters via the following property:
\begin{prop}\label{prop:1}
Let subsystem \eqref{eq:systP31} satisfy Assumption A3. Under the change of variable
\begin{equation}
	x = \frac{\mathrm{sgn}(2a_2^3-9a_3a_2a_1+27a_3^2a_0)}{3|a_3|}\sqrt{a_2^2-3a_3a_1}X -\frac{a_2}{3a_3},
\end{equation}
the coupled system \eqref{eq:systP31}-\eqref{eq:systP32} reduces to the canonical form
\begin{align}
	\frac{1}{\tilde{\epsilon}}\dot{X} &= 3X - X^3 + 2, \label{eq:systP31b}\\
\dot{\mathbf{y}} &= g(\mathbf{y},\tilde{\gamma} \dot{X}), \label{eq:systP32b}
\end{align}
with $\tilde{\epsilon}>0$ and $\tilde{\gamma}$ given by
\begin{equation}
    \tilde{\epsilon} = \epsilon \frac{3a_3a_1-a_2^2}{9a_3}\qquad \text{and}\qquad \tilde{\gamma} = \gamma\frac{\mathrm{sgn}(2a_2^3-9a_3a_2a_1+27a_3^2a_0)}{3|a_3|}\sqrt{a_2^2-3a_3a_1}.\label{eq:defgammaepstilde}
\end{equation}

\end{prop}
Since the proof consists of standard computations, namely the reduction of the cubic to depressed form followed by rescaling, we provide the details in Appendix~\ref{app:proof}. By also imposing Assumptions A1 on the fast subsystem \eqref{eq:systP32}, ensuring that it has a fold bifurcation at some critical value $b$ of the interaction component $\gamma\dot{x}$, the general safe-overshoot criterion of Result~\ref{res:1} can now be specialized to the following corollary:

\begin{corr}\label{corr:1}
Let the fast subsystem \eqref{eq:systP32} satisfy A1, and the slow subsystem \eqref{eq:systP31} satisfy A3. Let $x(t)$ be the unique heteroclinic trajectory such that
	\begin{equation}
		\lim_{t\to -\infty}x(t)=x_d,\qquad \lim_{t\to +\infty}x(t)=x_s,
	\end{equation}
and let $\tilde{\epsilon}$, $\tilde{\gamma}$ defined by \eqref{eq:defgammaepstilde}. Finite-time overshoot of the fast subsystem's fold threshold $b$ occurs if and only if  $b,\tilde{\gamma}>0$ and
\begin{equation}
	4\tilde{\gamma}\tilde{\epsilon}-b> 0.
\end{equation}
For sufficiently small $\epsilon$, such an overshoot occurring near the fast subsystem's fold bifurcation (in the sense of Remark \ref{rem:vicinityb}) is safe whenever 

\begin{equation}
\frac{F}{4} \sqrt{\frac{1}{6\tilde{\epsilon}^3\tilde{\gamma}}}\left(4\tilde{\gamma}\tilde{\epsilon}-b\right) \leq 1 + o(1),\label{eq:limsafeoverder}
\end{equation}
where $F = \alpha_0 \sqrt{2\kappa}$, and $o(1)$ denotes a quantity that vanishes as $\epsilon \to 0$.	
\end{corr}
To justify this corollary, we specialize Result~\ref{res:1} to the canonical form obtained in Property~\ref{prop:1}. This requires determining the maximum of the interaction component along the slow heteroclinic trajectory, and evaluating the quantity $S$ defined in Result~\ref{res:1}.

First, under Assumption A3, we rewrite system \eqref{eq:systP31}-\eqref{eq:systP32} via Property \ref{prop:1}, with $\tilde{\epsilon}$ and $\tilde{\gamma}$ defined by \eqref{eq:defgammaepstilde}. In the context of Result \ref{res:1}, we set $i=1$, and we have that $\gamma\epsilon c(X,\epsilon t)=\gamma\dot{X}$ which, in this case, has no explicit time dependency. Along the heteroclinic trajectory of the slow subsystem (for which existence and uniqueness follow immediately from the one-dimensional autonomous dynamics of the slow subsystem and assumption A3), $\dot{X}$ is strictly positive except at its endpoints, where it vanishes, and attains a unique non degenerate maximum in the interior of the trajectory. Therefore, under the sign conventions of assumption A1, a finite time overshoot requires the positivity of both $\tilde{\gamma}$ and $b$. The former ensures that the interaction component $\tilde{\gamma}\dot{X}$ has a regular maximum, and the latter that the overshoot, if it occurs, is finite in time.

In the case where an overshoot occurs, we may look for the maximum of the interaction component $\tilde{\gamma}\dot{X}$ along the slow subsystem's trajectory. This occurs at the maximum of $\dot{X}$, requiring $\ddot{X}=0$. Noting this maximum $\dot{X}_{\max}$, it translates into
\begin{equation}
	3(1-X^2)\dot{X}_{\max}=0,
\end{equation}
giving either $\dot{X}_{\max}=0$, corresponding to equilibria, or $X=\pm 1$. Since $X=-1$ corresponds to an equilibrium, the relevant interior maximum occurs for $X=1$. Hence, evaluating $\dot{X}$ for $X=1$ gives the value of $\dot{X}_{\max}$:
\begin{equation}
	\dot{X}_{\max} = \tilde{\epsilon}\left(3-1 + 2\right)= 4\tilde{\epsilon},
\end{equation}
so that $c_{\max}=4$ in the formalism of Result \ref{res:1}, and a finite time overshoot occurs if and only if $b,\tilde{\gamma}>0$ and
\begin{equation}
	4\tilde{\gamma}\tilde{\epsilon}-b>0.
\end{equation}
For sufficiently small $\epsilon$, and for such an overshoot occurring in the vicinity of the fast subsystem's fold bifurcation, we can use Result \ref{res:1}  to determine if the overshoot is safe. To do so, we evaluate the quantity $S$ given by equation \eqref{eq:S}. In the present context, we have that $f(X) = 3X-X^3+2$ and $c(X)=f(X)$, giving 
\begin{align}
    S &= \bigl[f^2\partial_1^2f+(\partial_1f)^2f](X(t_{\max}),\epsilon t_{\max}),\\
    &= -96.
\end{align}
Substituting this value of $S$, as well as $i=1$ and $c_{\max}=4$ into Result~\ref{res:1}, yields the claimed criterion \eqref{eq:limsafeoverder} for safe overshoot.

\begin{rem}\label{rem:vicinityb}
In Corollary~\ref{corr:1}, an overshoot is said to occur ``near the fast subsystem's fold bifurcation'' if the scaling regime introduced in the justification of Result~\ref{res:1}, adapted to the context of Corollary~\ref{corr:1}, holds:
\begin{enumerate}
	\item as the interacting component
$\gamma \dot{x}$ approaches the critical value $b$, the
fast variable remains sufficiently close to the attracting equilibrium branch connected to the fold with, in particular,
\begin{equation}
    \mathbf{w}_0^{\top}\left(\mathbf{y}(t) - \mathbf{y}_b\right) = O\!\left(\epsilon^{\frac{3}{4}}\right),
\end{equation}
	\item the overshoot magnitude is sufficiently small, satisfying
\begin{equation}
    4\tilde{\gamma}\tilde{\epsilon}- b
      = O\!\left(\epsilon^{\frac{3}{2}}\right).
\end{equation}
\end{enumerate}
\end{rem} 

Hence, Corollary~\ref{corr:1} provides a criterion equivalent to Result~\ref{res:1} in the simpler setting where the slow-component dynamics are governed by a scalar cubic polynomial and the coupling to the fast subsystem enters through a time derivative. 

\subsection{Application to example systems}\label{sec:Appexamples}
\begin{figure}[t!]
    \centering
    \includegraphics[scale=1.2]{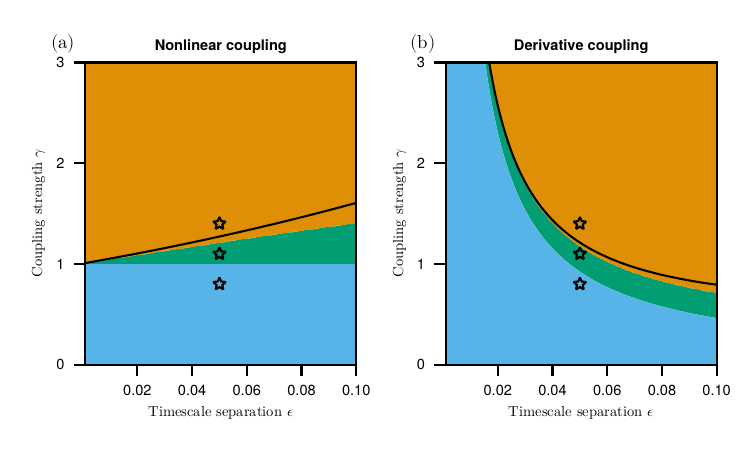}
    \caption{\emph{Overshoot regimes of the two example systems defined by \eqref{eq:systexx}-\eqref{eq:systexy} under the slowly varying forcing \eqref{eq:muillustr}. No-overshoot (blue), safe-overshoot (green), and unsafe-overshoot (orange) regions are shown as a function of timescale separation $\epsilon$ and coupling strength $\gamma$. \textnormal{(a)} Nonlinear coupling via $c_1(x)$. \textnormal{(b)} Derivative coupling via $c_2(x)$. The solid black line shows the approximation of the boundary between safe and unsafe overshoot provided by Result~\ref{res:1}. Stars indicate the parameter combinations used in Fig.~\ref{fig:Fig1}.}}
    \label{fig:Fig2}
\end{figure}

We now test the predictions of both Result~\ref{res:1} and Corollary~\ref{corr:1} against numerical experiments similar to those presented in Section~\ref{sec:illustr}. In these experiments, the external forcing is slowly increased using equation~\eqref{eq:muillustr} for both example systems defined by \eqref{eq:systexx}-\eqref{eq:systexy}, while varying the timescale separation $\epsilon$ and the coupling strength $\gamma$. Figure~\ref{fig:Fig2} summarizes the outcomes of these experiments. For both coupling configurations, three distinct regions can be distinguished in the $(\epsilon,\gamma)$ parameter space: a blue region where the fold of the fast subsystem is never crossed, an orange region where overshoot occurs and leads to tipping of the fast subsystem, and an intermediate green region in which the overshoot is safe.

We first apply Result~\ref{res:1} to approximate the boundary between safe and unsafe overshoot. In the case of nonlinear coupling (Figure~\ref{fig:Fig2}a), where the interaction is mediated by the nonlinear function of the slow variable $c_1(x)$, we set $i = 0$ and $c(\mathbf{x},\epsilon t) = c_1(x)$. In contrast, in the case of derivative coupling (Figure~\ref{fig:Fig2}b), in which the interaction is via the time derivative $\dot{x}$, we set $i = 1$ and $c(\mathbf{x},\epsilon t) = c_2(x,\epsilon t) = \bigl(3x - x^3 + \mu(\epsilon t)\bigr)$. Within this framework, we obtain $\alpha_0 = 1$ and $\kappa = 2$ in Assumption~A1, yielding $F = \alpha_0 \sqrt{2\kappa} = 2$. The corresponding expressions for $S$, given by equation~\eqref{eq:S}, are computed accordingly, with $\epsilon t_{\max}$ evaluated numerically.

In both coupling configurations, we find that the approximation provided by Result~\ref{res:1} (solid black line in Figure~\ref{fig:Fig2}) captures the boundary between safe and unsafe overshoot well. As expected from the asymptotic nature of Result \ref{res:1}, the agreement improves with increasing timescale separation, that is, as $\epsilon$ decreases. Importantly, the result captures the qualitatively different influence of the two types of coupling on the behaviour of this boundary, or equivalently, on the critical coupling strength $\gamma_c$ above which the overshoot becomes unsafe.

This difference becomes explicit in the asymptotic limit $\epsilon \to 0$, where this critical coupling strength is determined solely by the condition for the occurrence of an overshoot, as given by the leading-order form of criterion~\eqref{eq:safecritgen}. For nonlinear coupling ($i = 0$), the leading-order form of criterion~\eqref{eq:safecritgen} yields
\begin{equation}
	\gamma_c \;=\; \frac{b}{c_{\max}} \;+\; O(\epsilon),
\end{equation}
so that the critical coupling strength approaches a constant value linearly in $\epsilon$. In contrast, for derivative coupling ($i = 1$), the criterion yields
\begin{equation}
	\gamma_c \;=\; \frac{b}{c_{\max}}\,\epsilon^{-1} \;+\; O(1),
\end{equation}
implying that the critical coupling strength grows without bound, scaling as $\gamma_c \sim \epsilon^{-1}$. This reflects the fact that derivative coupling is intrinsically scaled by the slow timescale, whereas nonlinear coupling is not, leading to fundamentally different asymptotic behaviour in regimes of strong timescale separation.

We next consider the application of Corollary~\ref{corr:1}. In the case of system~\eqref{eq:systexx}-\eqref{eq:systexy} with derivative coupling, Assumption~A3 is satisfied at $\mu = \pm 2$. In what follows, we consider $\mu = 2$, which corresponds to the tipping point that is relevant to these experiments. At this fixed value of the forcing, the system already matches the form~\eqref{eq:systP31b}--\eqref{eq:systP32b}, with $\tilde{\epsilon} = \epsilon$ and $\tilde{\gamma} = \gamma$. Accordingly, for a sufficiently slow passage through the fold bifurcation of the slow subsystem (i.e., for sufficiently small $r$ in equation~\eqref{eq:muillustr}), we expect the slow transition to occur in a neighbourhood of $\mu = 2$, close to the heteroclinic trajectory invoked in Corollary~\ref{corr:1}. Thus, although the slow dynamics are assumed autonomous in Corollary~\ref{corr:1}, it can be applied to transient experiments of the type considered here, with its predictions interpreted as representative of the limit in which the forcing rate tends to zero.

\begin{figure}[b!]
    \centering
    \includegraphics[scale=1.2]{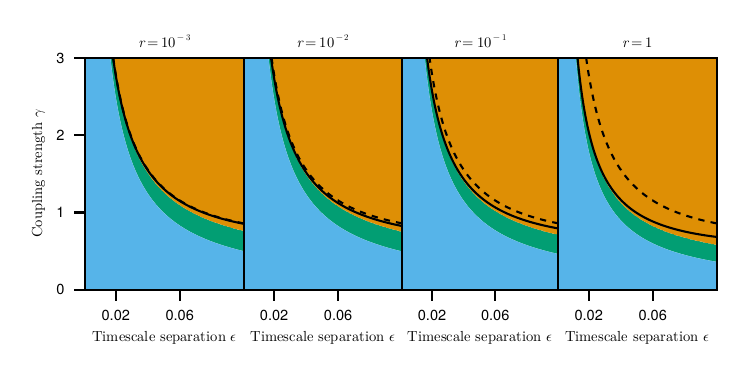}
    \caption{\emph{Overshoot regimes of the example system defined by \eqref{eq:systexx}-\eqref{eq:systexy} in the case of derivative coupling via $c_2(x)$, for different values of the forcing rate parameter $r$. No-overshoot (blue), safe-overshoot (green), and unsafe-overshoot (orange) regions are shown as a function of timescale separation $\epsilon$ and coupling strength $\gamma$. The solid and dashed black lines show the approximation of the boundary between safe and unsafe overshoot provided by Result~\ref{res:1} and Corollary~\ref{corr:1}, respectively.}}
    \label{fig:Fig3}
\end{figure}

This behaviour is illustrated in Figure~\ref{fig:Fig3}, which shows the outcomes of such transient experiments in the $(\epsilon,\gamma)$ parameter space for increasing values of the rate parameter $r = \{10^{-3}, 10^{-2}, 10^{-1}, 1\}$ in equation~\eqref{eq:muillustr}. The predicted boundary between safe and unsafe overshoot is overlaid for both Result~\ref{res:1} (black solid line) and Corollary~\ref{corr:1} (black dashed line). Result~\ref{res:1} remains accurate across all rates considered, whereas the predictions of Corollary~\ref{corr:1} become progressively less accurate as the rate increases. This discrepancy arises because different values of the rate $r$ lead to different values of both the slow subsystem state and the time at which the overshoot peak occurs, that is, $x(t_{\max})$ and $\epsilon t_{\max}$ in Result~\ref{res:1}. While these quantities enter explicitly into the criterion of Result~\ref{res:1}, Corollary~\ref{corr:1} effectively evaluates them at the critical value of the forcing parameter. As the rate decreases, however, the two predictions approach one another and become nearly indistinguishable, demonstrating that Corollary~\ref{corr:1} is reliable in the small-rate regime, where non-autonomous effects become negligible.

\section{Application to interacting climate tipping elements}\label{sec:Applications}
In this section, we illustrate the applicability of the theoretical results developed in Section~\ref{sec:Crit} by analyzing two conceptual slow-fast systems representing interacting climate tipping elements. In Section~\ref{sec:CessiVeg}, we consider the interacting Atlantic Meridional Overturning Circulation (AMOC) and Amazon rainforest, allowing to test the predictions of Result~\ref{res:1} under nonlinear interaction. In Section~\ref{sec:GISAMOC}, we examine a coupled AMOC-Greenland Ice Sheet (GIS) model in which a derivative coupling arises from the interaction with GIS meltwater fluxes which, together with the polynomial structure of the GIS model, enables the use of Corollary~\ref{corr:1}. 

To provide a physical interpretation and estimate an upper bound for the timescale separation $\epsilon$, we interpret variations in $\epsilon$ as representing changes in the tipping duration of the slow subsystem. Additional details on this interpretation, together with the selection of suitable values and ranges for other parameters, are provided in Appendix~\ref{app:estimation}. Finally, we primarily consider low rates of change in the external forcing, which allows us to focus on the consequences of interactions between tipping elements in the regime where the forcing rate itself does not dominate the slow dynamics. We nevertheless assess the robustness of our results using faster forcing rates in supplementary experiments.
\subsection{Interacting AMOC and Amazon rainforest}\label{sec:CessiVeg}

 The Atlantic Meridional Overturning Circulation (AMOC) is a major climate tipping element, affected by density gradients within the Atlantic Ocean. Its tipping dynamics are mostly understood in terms of the positive salt-advection feedback, which can render the circulation susceptible to abrupt transitions under sufficiently strong freshwater perturbations, as has been reproduced in models of various complexity \cite{stommel_thermohaline_1961,rahmstorf_thermohaline_2005,van_westen_physics-based_2024}. Here, we describe the evolution of the AMOC using a two-box model \cite{cessi_simple_1994}. This model captures the dynamics of the meridional temperature and salinity differences within the Atlantic, denoted by $x$ and $y$, respectively, and evolves on a nondimensional slow time $\tau$, where one unit corresponds to the diffusion timescale $\theta_\text{diff} =  180$~years. The governing equations are
\begin{equation}
\begin{aligned}
    x' &= -\alpha (x - 1) - xQ(x,y), \\
    y' &= F_0 - yQ(x,y),
\end{aligned}
\qquad
\text{where}
\qquad	
Q(x,y) = 1+\mu^2(x-y)^2,
\end{equation}
where the prime denotes differentiation with respect to the slow time $\tau$. Both the $x$- and $y$-equations include an advective contribution proportional to the AMOC strength $Q(x,y)$, and the forcing parameter $F_0=1.1$ represents the meridional freshwater-forcing difference in the reference state. Numerical values for all parameters are taken from \cite{dijkstra_nonlinear_2013} and listed in Table~\ref{tab:par} with a short description. In the reference configuration, the AMOC is bistable. However, as the freshwater forcing increases to the critical threshold $F_{\mathrm{c}} \approx 1.29$, the system exhibits a fold bifurcation. After this point, the strong AMOC state loses stability, leaving only a collapsed circulation stable.

 The Amazon rainforest is another climate tipping element, whose stability is shaped by interactions between many components, including vegetation, precipitation, and fire. To describe its dynamics, we use the conceptual model introduced in \cite{van_nes_tipping_2014}. This model captures the evolution of precipitation rate $P$ and tree-cover fraction $T$ over the Amazon basin, and evolves on a dimensional time expressed in years. To work consistently with nondimensional time, we treat this fast time $t$ as dimensionless, taking one unit to correspond to the vegetation adjustment timescale $\theta_{\text{veg}} = 1$ year. In this formulation, parameters retain their original numerical values, but their units adjust accordingly. The precipitation dynamics are given by
\begin{equation}
    \dot{P} = r_P\left(P_d + b_f\,\frac{T}{K} - P\right), \label{eq:prec}
\end{equation}
 where the dot denotes differentiation with respect to the fast time $t$. This equation represents a vegetation-precipitation feedback, in which precipitation relaxes toward an equilibrium value that increases with tree cover. In particular, $P_d$ is the background precipitation in the absence of trees, while $b_f$ quantifies the strength of the vegetation-precipitation feedback. The tree-cover dynamics are described by
\begin{equation}
\begin{aligned}
    \dot{T} &= \frac{P}{h_P + P}\, r_m\, T\left(1 - \frac{T}{K}\right)
             - m_A\, T\,\frac{h_A}{T + h_A}
             - m_f\, T\,\frac{h_f^\beta}{h_f^\beta + T^\beta}.
\end{aligned}
\end{equation}
 This equation consists of three different contributions. The first term represents logistic tree growth modulated by precipitation availability. The second term introduces an Allee effect, increasing mortality at low tree cover. Finally, the third term captures mortality from fires, represented by a saturating mortality function. The value of all parameters and their description are given in Table \ref{tab:par} with a short description (see also Appendix \ref{app:estimation} for estimation of $P_d$ and $b_f$). Under decreasing background precipitation $P_d$, the model displays different stable branches of equilibria: forest, savanna, and desert states organised by fold bifurcations occurring at critical precipitation levels. In what follows, we focus on the transition from the forest state to the savanna state occurring at the critical value $P_{d,\text{c}}\approx1.93$ mm$/$day.

 Several studies have found that a weakening AMOC causes a southward shift of the Inter-tropical Convergence Zone and altered rainfall patterns \cite{jackson_global_2015, parsons_influence_2014}, which affect the Amazon rainforest productivity. We therefore couple both models under the simplifying assumption that variations in the AMOC strength $Q$ linearly influence precipitation over the Amazon basin. Specifically, we add to the background precipitation a term proportional to the AMOC strength anomaly $\Delta Q$ relative to its value $Q_0 = 4.55$ in the reference AMOC state, rewriting equation \eqref{eq:prec} as follows:
\begin{equation}
    \dot{P} = r_P\left(P_d +\gamma\Delta Q+ b_f\,\frac{T}{K} - P\right),
    \qquad \text{where} \qquad
    \Delta Q = Q - Q_0.\label{eq:Pamaz}
\end{equation}
Here, $\gamma$ is a coupling strength that quantifies how changes in AMOC strength affect precipitation over the Amazon, with $\gamma>0$ implying that a weakening AMOC reduces precipitation. In this section, we will consider this case, as the case $\gamma<0$ will not yield overshoot of the Amazon system's fold bifurcation.  Expressed in terms of the fast, yearly timescale, and defining the reference timescale separation $\epsilon = \theta_\text{veg}/\theta_\text{diff}=1/180$, the interacting AMOC-Amazon system forms a slow-fast system reading as 
\begin{align}
	\frac{1}{\epsilon}\dot{x} &= -\alpha (x-1) - x(1+\mu^2(x-y)^2),\\
    \frac{1}{\epsilon}\dot{y} &= F_0 - y(1+\mu^2(x-y)^2) \label{eq:CessiVegy},\\
    \dot{P} &= r_P\left(P_d+\gamma\Delta Q(x,y)+b_f\frac{T}{K}-P\right),\\
    \dot{T} &= \frac{P}{h_P+P}r_mT\left(1-\frac{T}{K}\right) - m_AT\frac{h_A}{T+h_A}-m_fT\frac{h_f^\beta}{h_f^\beta+T^\beta}.
\end{align}
Starting from the reference state with strong AMOC and the Amazon rainforest model in a forest state (see Table \ref{tab:init}), we perform transient experiments in which we slowly increase the freshwater forcing on the AMOC model. In particular, we replace the reference forcing term $F_0$ in equation \eqref{eq:CessiVegy} by a time dependent freshwater forcing $F(t)$ as follows
\begin{equation}
	F(t) = F_0+r \epsilon t, \label{eq:Fvar}
\end{equation}
where  $r = 10^{-4}$ sets the rate of change in the slow-time $\epsilon t$. The experiments are conducted for $t\in[0,6\times 10^5/\theta_{\text{veg}}]$, corresponding to $6\times 10^5$ model years, over the relevant parameter range for the coupling strength \(\gamma \in [0.0, 0.88]\), and for timescale separations up to \(\epsilon = 0.045\), the latter corresponding to an AMOC collapse unfolding on a timescale of approximately 50 years. Details on how these parameter ranges are determined are provided in Appendix~\ref{app:estimation}.

\begin{figure}[t!]
    \centering
    \includegraphics[scale=1.2]{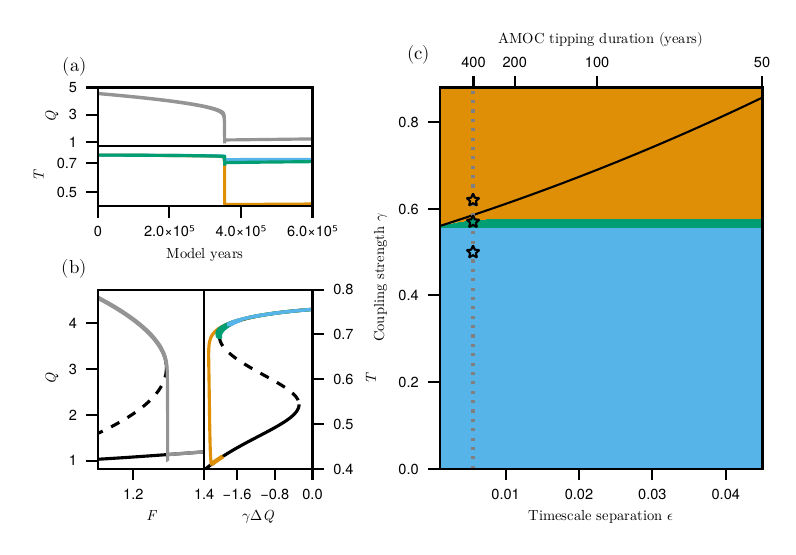}
    \caption{\emph{Coupled AMOC-Amazon overshoot dynamics under the slowly varying forcing \eqref{eq:Fvar}. \textnormal{(a)} Trajectories of AMOC strength $Q$ (top) and Amazon tree cover $T$ (bottom) for three values of the coupling strength $\gamma = 0.50$, $0.57$, and $0.62$ (blue, green, and orange, respectively). \textnormal{(b)} The same trajectories as in \textnormal{(a)}, shown on the bifurcation diagrams of the AMOC (left) and Amazon (right), with stable (solid black) and unstable (dashed black) equilibria. \textnormal{(c)} Overshoot regimes of the coupled AMOC-Amazon system. No-overshoot (blue), safe-overshoot (green), and unsafe-overshoot (orange) regions are shown as a function of timescale separation $\epsilon$ and coupling strength $\gamma$. The solid black line shows the approximation of the boundary between safe and unsafe overshoot provided by Result~\ref{res:1}. Stars indicate the parameter combinations shown in (a-b). The vertical dotted grey line denotes the reference value of $\epsilon$.}}
    \label{fig:Fig4}
\end{figure}

The results across the $(\epsilon,\gamma)$ parameter space are summarized in Figure~\ref{fig:Fig4}c, with the corresponding AMOC tipping durations indicated along the upper horizontal axis and representative trajectories shown in Figure~\ref{fig:Fig4}.a-b. For sufficiently low values of the coupling strength (i.e., $\gamma < 0.55$), the interaction component $\gamma\Delta Q$ never overshoots the fold threshold of the Amazon model (blue in Figure~\ref{fig:Fig4}a-c). Nevertheless, in all cases the AMOC strength in our simulations briefly dips below its asymptotic collapsed value, reaches a temporary minimum, and then relaxes upward (grey curves in Figure~\ref{fig:Fig4}a and \ref{fig:Fig4}.b), a behaviour that was also found in another conceptual model \cite{sinet_amoc_2023}. Consequently, for higher coupling strengths (i.e., $\gamma \ge 0.55$), the interaction term can cross the fold threshold for a finite time, producing overshoots that may be safe or unsafe (green and orange regions in Figure~\ref{fig:Fig4}a-c, respectively). Taken together, our model results suggest that safe overshoots may not be a particularly robust feature of the coupled AMOC-Amazon system, as the corresponding region (in green) only exists within a narrow band of the plausible coupling strength range. However, when the forcing rate $r$ is increased, the safe-overshoot region extends to higher values of $\gamma$ (see Figure \ref{fig:FigS1}). This occurs because, although the overshoot magnitude remains similar, the AMOC spends less time near its minimum, leading to a comparatively shorter overshoot duration.

Next, we compare these numerical results with the criterion given in Result~\ref{res:1}. In the present application, the extremum of the interaction component $\gamma \Delta Q$ corresponds to a minimum rather than a maximum, which can be reduced to Assumptions~A1-A2 by an appropriate reparametrization. We determine numerically the time at which the interaction component attains its extremal value along the slow subsystem trajectory, and evaluate the coefficient $F$ appearing in criterion~\eqref{eq:safecritgen} numerically using equation~\eqref{eq:F}, and the coefficient $S$ directly from expression~\eqref{eq:Sdef}. At the lowest values of $\epsilon$, the boundary between safe- and unsafe-overshoot regimes is well approximated by Result~\ref{res:1}, shown as the black curve in Figure~\ref{fig:Fig4}c. However, we find that the predictive skill of Result \ref{res:1} depends critically on both the amplitude and the temporal structure of the AMOC anomaly. As $\epsilon$ increases, the analytical approximation diverges rapidly from the numerical results, systematically predicting larger values of $\gamma$ for the onset of unsafe overshoot than those seen in the simulations. This is also found in the cases where the rate of forcing $r$ is higher (see Figure \ref{fig:FigS1}). It occurs because, at these large values of $\gamma$, the overshoot is not small enough to be well approximated by a local quadratic expansion around its extremum, owing to the strongly asymmetric variation in the AMOC strength anomaly. This limitation is expected, but becomes especially relevant when the overshoot trajectory of the interacting component is strongly non-parabolic, as is the case here.

\subsection{Interacting Greenland Ice Sheet and AMOC}\label{sec:GISAMOC}

Recent literature suggests that the Greenland Ice Sheet (GIS), also a key climate tipping element, is already at risk of crossing its tipping point under present-day conditions \cite{armstrong_mckay_exceeding_2022}. Owing to its inherently slow, millennial-scale response, the GIS is considered here as the slow subsystem. To represent its large-scale behaviour, we adopt the approach introduced in the reduced model SURFER \cite{martinez_montero_surfer_2022}, which describes the evolution of the GIS ice-volume fraction $V$ as 
\begin{equation}
    V' = -V^3 + a_2 V^2 + a_1 V + s\, \delta T_{0} + a_0,
\end{equation}
where the prime denotes differentiation with respect to the nondimensional slow time~$\tau$, obtained by scaling time with the characteristic GIS melting timescale, taken as $\theta_{\text{melt}}=470$~years. Although the reduced model does not explicitly resolve key ice-sheet processes (e.g., the melt-elevation feedback), its parameters are calibrated against the output of a comprehensive ice-sheet model \cite{robinson_multistability_2012}. The melting timescale has been selected to match the transient response of the GIS to sustained warming, and the coefficients $a_{1,2}$ and $s$ have been selected to reproduce part of the GIS bifurcation structure. The latter are defined in terms of $V_{+,-}$ and $T_{+,-}$, i.e., the GIS ice-volume fraction and the temperature anomaly relative to pre-industrial conditions at the fold bifurcations associated with GIS melting and regrowth (freezing), respectively (see relations in \cite{martinez_montero_surfer_2022}). In this model, $\delta T_{0} = 1.1^\circ\mathrm{C}$ represents the reference present-day-like global-mean surface warming relative to pre-industrial conditions, for which the GIS is bistable. As this parameter increases to the critical value $\delta T_{c} = 1.52^\circ\mathrm{C}$, the model undergoes a fold bifurcation, beyond which the stable present-day GIS state disappears.

It has been shown that meltwater released by the GIS, by freshening the North Atlantic, negatively impacts the stability of the AMOC \cite{sinet_meltwater_2025,jackson_global_2015,li_global_2023}. In addition to its sensitivity to the magnitude of this meltwater flux, the AMOC is also known to respond to the rate at which the meltwater flux varies, with rate-induced effects demonstrated in both conceptual \cite{sinet_amoc_2024,klose_rate-induced_2024, chapman_tipping_2024} and comprehensive \cite{lohmann_risk_2021} models. The prominence of such rate-dependent responses, of which safe overshoots are an example, motivates coupling the GIS to the AMOC to explore whether safe overshoots can occur.

Therefore, we couple the GIS model to the AMOC model presented in the previous section, this time considering the AMOC model's nondimensional time as the fast time~$t$. As the added meltwater flux is proportional to the ice loss from the GIS, it is reflected in the AMOC model as an additional freshwater contribution proportional to $\dot{V}$, the time derivative of the GIS ice-volume fraction. Expressing the coupled system on the fast timescale, and defining the reference timescale separation $\epsilon = \theta_\text{diff} / \theta_\text{melt}$, the full coupled model reads
\begin{align}
    \frac{1}{\epsilon}\dot{V} &= -V^3 + a_2 V^2 + a_1 V + s \delta T_{0} + a_0, \label{eq:GIScoupled}\\
    \dot{x} &= -\alpha (x-1) - x(1+\mu^2(x-y)^2),\\
    \dot{y} &= F_0 - \gamma\dot{V} - y(1+\mu^2(x-y)^2), 
\end{align}
where the dot denotes derivatives with respect to the fast time $t$. In this coupled model, the reference value of $\gamma$ can be evaluated as $\gamma \approx 3.8$, given directly by how much a variation in the GIS ice volume affects the density contrast between the two AMOC model boxes (see Appendix \ref{app:estimation}).

From the reference state with both the GIS and AMOC on their present-day-like stable branch (see Table \ref{tab:init}), we investigate the response of the coupled AMOC-GIS system under transient simulations in which the temperature forcing applied to the GIS is gradually increased. The reference forcing term $\delta T_0$ in equation~\eqref{eq:GIScoupled} is replaced by a time-dependent function,
\begin{equation}
    \delta T(t) = \delta T_{0} + r\epsilon t, \label{eq:vardT}
\end{equation}
where $r$ sets the rate of warming in the slow time $\epsilon t$, here chosen as $r = 10^{-4}$ $^\circ$C. The experiments are conducted for $t\in[0,10^6/\theta_{\text{diff}}]$, corresponding to $10^6$ model years. We explore timescale separations up to $\epsilon = 3.3$, corresponding to a GIS collapse occurring on a timescale of approximately $1000$ years. While this value of $\epsilon$ may appear large in the context of classical slow-fast systems, it remains sufficiently small to ensure a clear separation between the timescales of GIS and AMOC tipping, with the GIS response remaining substantially slower than that of the AMOC for all values of $\epsilon$ considered here. Finally, we consider coupling strengths in the range $\gamma \in [0, 10]$, encompassing the physically estimated coupling value $\gamma \approx 3.8$ and allowing us to investigate how the system's behaviour depends on the strength of the interaction. More details are provided in Appendix~\ref{app:estimation}. 

\begin{figure}[t!]
    \centering
    \includegraphics[scale=1.2]{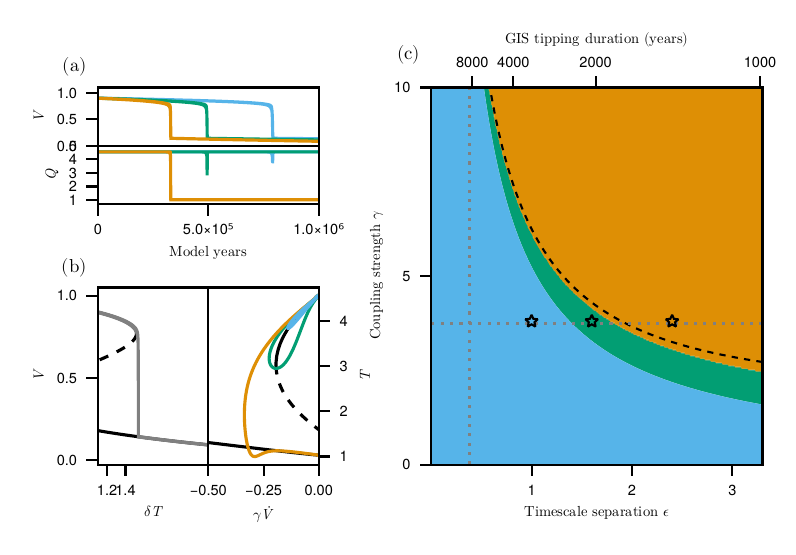}
    \caption{\emph{Coupled GIS-AMOC overshoot dynamics under the slowly varying forcing \eqref{eq:vardT}. \textnormal{(a)} Trajectories of GIS ice-volume fraction $V$ (top) and AMOC strength $Q$ (bottom) for three values of the timescale separation $\epsilon = 1.0$, $1.6$, and $2.4$ (blue, green, and orange, respectively). \textnormal{(b)} The same trajectories as in \textnormal{(a)}, shown on the bifurcation diagrams of the GIS (left) and AMOC (right), with stable (solid black) and unstable (dashed black) equilibria. \textnormal{(c)} Overshoot regimes of xthe coupled GIS-AMOC system. No-overshoot (blue), safe-overshoot (green), and unsafe-overshoot (orange) regions are shown as a function of timescale separation $\epsilon$ and coupling strength $\gamma$. The dashed black line shows the approximation of the boundary between safe and unsafe overshoot provided by Corollary~\ref{corr:1}. Stars indicate the parameter combinations shown in (a-b). The vertical and horizontal dotted grey lines denote the reference values of $\epsilon$ and $\gamma$, respectively.}} 
    \label{fig:Fig5}
\end{figure}

The results are displayed across the $(\epsilon,\gamma)$ parameter space in Figure~\ref{fig:Fig5}c, with corresponding GIS tipping durations indicated along the top axis and illustrative trajectories shown in Figure~\ref{fig:Fig5}a-b. The qualitative behaviour closely mirrors that of the generic system with derivative coupling in Figure~\ref{fig:Fig2}: overshoots occur only when both the timescale separation and the coupling strength are sufficiently large, since the overshoot magnitude is proportional to both parameters. A band of safe overshoot also emerges, shifting toward lower coupling strengths as $\epsilon$ increases, that is, as the GIS collapses more rapidly. The reference values of $\epsilon$ and $\gamma$ (intersection of the dashed grey lines) lie well within the no-overshoot region, indicating that present-day parameter estimates place the system firmly in a non-overshooting regime. At the same time, for the reference coupling strength, the diagram shows that safe overshoot becomes possible for GIS tipping durations between about 2355 and 1810 years, a roughly 545-year window that is non-negligible relative to millennial timescales expected for a fast GIS collapse.

As the slow subsystem's dynamical equation is a cubic polynomial and the interaction enters through its time derivative, we use Corollary~\ref{corr:1} as an approximation for the boundary between safe and unsafe overshoots in parameter space. Up to an appropriate reparametrization, Assumptions~A1 and~A3 are satisfied, and all that is needed is to evaluate the coefficient $F$ appearing in criterion~\eqref{eq:safecritgen}, for example, through numerical continuation using equation~\eqref{eq:F}. The numerically determined boundary between safe and unsafe overshoots agrees well with the approximation obtained from Corollary~\ref{corr:1} (dashed black curve). Although the analytical boundary slightly overestimates the coupling strength required at larger timescale separations, the discrepancy remains modest across the explored parameter range. However, our experiments do not strictly satisfy the assumptions of the Corollary \ref{corr:1}, as they involve a non-autonomous slow passage through a fold rather than the heteroclinic trajectory of an autonomous system. For the small forcing rate used here, the non-autonomous trajectory stays close to the corresponding heteroclinic orbit (see Figure \ref{fig:Fig5}b, left), so the prediction of the corollary remains accurate. As the forcing rate increases, this approximation breaks down: the slow subsystem no longer undergoes a genuine slow passage through the fold, and the boundary between safe and unsafe overshoot is less well captured (see Fig. \ref{fig:FigS2}). In these cases, however, Result \ref{res:1} remains applicable, since it does not rely on this hypothesis (see Fig. \ref{fig:FigS2}).

\section{Conclusion}\label{sec:Conclusion}
In this paper, we extended the theoretical framework for safe-overshoot phenomena to interacting systems, motivated by their potential importance in the context of coupled tipping elements and cascading phenomena. For this, we considered slow-fast systems interacting only in one direction, from the slow to the fast component, and coupled through nonlinear functions or time derivatives.

We formulated a general result that applies to such systems in which the fast component has a fold bifurcation that is overshot for a finite time due to its interaction with the slow subsystem. In the case of interactions via both general nonlinear functions and time derivatives, this result provides a criterion for distinguishing safe from unsafe overshoots. As was the case in the context of isolated systems \cite{ritchie_inverse-square_2019}, this criterion is valid in the case of strong timescale separation and small overshoot. Within a framework of interacting tipping systems, our criterion is expressed in terms of the timescale separation and coupling strength in an inverse-square law similar to \cite{ritchie_inverse-square_2019}. Next, we established a corollary for systems in which the slow component follows a third-degree polynomial, a standard archetype for tipping phenomena \cite{wunderling_interacting_2021}, and in which coupling occurs via a time derivative, a natural choice in conceptual representations of the interacting ice sheets and AMOC \cite{sinet_amoc_2023,sinet_amoc_2024,klose_rate-induced_2024}. This corollary applies to such systems in the autonomous case, but can also be used to explore non-autonomous regimes such as a slow passage through the slow subsystem's fold bifurcation.

These results were illustrated using simple illustrative examples and two models of coupled tipping elements: an AMOC-Amazon rainforest model interacting via AMOC strength (nonlinear coupling), and a GIS-AMOC model interacting via meltwater fluxes (derivative coupling). Together, these case studies suggest that our results are practically useful, while also clarifying their limitations, largely mirroring those of the inverse-square law for isolated systems \cite{ritchie_inverse-square_2019}. On the one hand, in the coupled AMOC-Amazon system, we find that the safe-overshoot regime is confined to a narrow band of the plausible parameter range, and is therefore not a particularly robust characteristic of the model. However, comparison with the analytical criterion (Result \ref{res:1}) demonstrates good predictive skill only in the small-overshoot regime, and rapidly overestimates the safe-overshoot region for weaker timescale separation (larger~$\epsilon$) due to strongly non-parabolic and asymmetric overshoot dynamics. On the other hand, in the coupled AMOC-GIS case, the numerical simulations and analytical approximations show good overall agreement across the explored parameter space. Numerically, overshoots occur only when both the coupling strength and the timescale separation are sufficiently large, and a band of safe overshoot emerges that shifts toward lower~$\gamma$ as $\epsilon$ increases (i.e., as the GIS collapses more rapidly). The boundary between safe and unsafe overshoots is well captured by the approximation from Corollary~\ref{corr:1} for slow forcing rates, while Result~\ref{res:1} remains accurate at higher forcing rates, allowing us to conclude, for example, that for physically relevant GIS-AMOC coupling, safe overshoot may occur over a GIS tipping-duration window of roughly 545~years. Overall, these case studies show that our criteria offer a useful first-order diagnostic of overshoot risk in coupled systems, while making clear when their assumptions break down and numerical investigation is required.

This work points to several directions for future research. First, coupled systems with leading-following (also referred to as upstream–downstream in \cite{ashwin_early_2025}) structures like those considered here have been studied previously. An interesting next step is to investigate how the geometry of their underlying bifurcation structure \cite{sinet_approximating_2025} shapes the existence and extent of safe-overshoot regimes. In the same context, it would be valuable to explore how a criterion for safe overshoot can be used in conjunction with early-warning indicators \cite{ashwin_early_2025}. Second, the inverse-square law for isolated systems has seen multiple extensions, including to stochastic settings \cite{ritchie_inverse-square_2019}. Extending such ideas to coupled systems by introducing noise in either the slow or fast subsystem would enable a probabilistic notion of safe- versus unsafe-overshoot windows that accounts for noise-induced transitions. Likewise, uncertainty quantification has been applied to the inverse-square law for isolated systems \cite{lux-gottschalk_uncertainty_2025}. Extending these methods to the coupled context would allow safe-overshoot windows and their boundaries to be expressed with calibrated uncertainty. Finally, it would be valuable to consider more complex and realistic settings. On the one hand, this includes extending the theoretical approach beyond the minimal two-component, unilateral setup to more general interaction structures, including feedback loops, additional components, and possibly spatially extended dynamics. On the other hand, it involves assessing how the proposed criteria perform in more comprehensive models such as Earth System Models, where multiple adjustment timescales, heterogeneity, and internal variability complicate, among many other things, the notion of timescale separation.

To conclude, we have extended the inverse-square-law formalism of \cite{ritchie_inverse-square_2019} to a relevant class of interacting slow-fast systems, providing a simple, interpretable criterion for when an overshoot of a fold threshold is likely to remain safe. In doing so, we offer a first-order way to map overshoot risk across coupling strengths and timescale separations. More broadly, by shifting attention from fixed tipping points to transient overshoots, our results provide a complementary step toward assessing cascading tipping risk in coupled systems, up to and including the Earth system.

\vskip6pt

\enlargethispage{20pt}

\paragraph{Acknowledgements} This project has received funding from the Dutch Research Council (NWO) through the NWO-Vici
project ``Interacting tipping elements: When does tipping cause
tipping'' (Project No. VI.C.202.081). P.D.L.R acknowledges funding from the European Union's Horizon Europe research and innovation program under the grant agreement No. 101137601 (ClimTip), as well as funding by the UK Advanced Research and Invention Agency (ARIA) via the project ``AdvanTip''.

\paragraph{Data availability} All codes used to produce data and figures are publicly available at \href{https://github.com/sachasinet/CascAnalytics}{https://github.com/sachasinet/CascAnalytics}.

\appendix

\section{Proof of Property \ref{prop:1}}\label{app:proof}
In this appendix, we provide the proof of Property \ref{prop:1}, consisting in the reduction of the cubic in equation \eqref{eq:systP31} to depressed form followed by rescaling. In the dynamical equation
\begin{equation}
	\frac{1}{\epsilon}\dot{x} = a_3x^3+a_2x^2+a_1x+a_0,
\end{equation}
the r.h.s.\ can be reduced to a depressed cubic (i.e. a cubic without a square term) via the change of variable
\begin{equation}
	x = \xi-\frac{a_2}{3a_3},
\end{equation}
allowing us to rewrite it, after dividing the equation by $a_3$, as
\begin{equation}
	\frac{1}{a_3\epsilon}\dot{\xi} = \xi^3 + p\xi + q, \label{eq:xi}
\end{equation}
where $p$ and $q$ are given by
\begin{align}
	p &= \frac{3a_3a_1-a_2^2}{3a_3^2},\\ 
	q &= \frac{2a_2^3-9a_3a_2a_1+27a_3^2a_0}{27a_3^3}.
\end{align}
By Assumption A3, we have that $4p^3+27q^2=0$ and $p\neq 0$, such that $p<0$. Hence, we may introduce the rescaling
\begin{equation}
	\xi = -\operatorname{sgn}(q)\sqrt{-\tfrac{p}{3}}\,X.
\end{equation}
Substituting this expression into equation \eqref{eq:xi}, and dividing it by $\mathrm{sgn}(q)(-\frac{p}{3})^\frac{3}{2}$,  the equation becomes
\begin{equation}
	\frac{1}{\epsilon}\frac{3}{a_3p}\dot{X} 
	= 3X - X^3 + \frac{q}{\operatorname{sgn}(q)} \Bigl(-\tfrac{3}{p}\Bigr)^{3/2}.
\end{equation} 
It remains to compute the constant term. The relation $4p^3+27q^2=0$ implies that $|q| = 2\bigl(-\tfrac{p}{3}\bigr)^{3/2}$, so that
\begin{align}
	\frac{q}{\operatorname{sgn}(q)} \Bigl(-\tfrac{3}{p}\Bigr)^{3/2} &= |q|\Bigl(-\tfrac{3}{p}\Bigr)^{3/2},\\
					&=2,
\end{align}
giving the final form
\begin{equation}
	\frac{1}{\tilde\epsilon}\dot{X} = 3X - X^3 + 2,
\end{equation}
with $\tilde \epsilon = \epsilon\frac{3a_3a_1-a_2^2}{9a_3}>0$, as claimed. Since $\dot x=-\operatorname{sgn}(q)\sqrt{-\tfrac{p}{3}}\,\dot X$, the slow equation reads
\begin{equation}
	\dot{\mathbf y}=g\bigl(\mathbf y,\tilde\gamma\,\dot X\bigr),
\end{equation}
with
\begin{align}
	\tilde\gamma &= -\gamma\,\operatorname{sgn}(q)\sqrt{-\tfrac{p}{3}},\\
	&= \gamma\frac{\operatorname{sgn}(2a_2^3-9a_3a_2a_1+27a_3^2a_0)}{3|a_3|}\sqrt{a_2^2-3a_3a_1},
\end{align}
concluding the proof.

\section{Parameter estimations}\label{app:estimation}

In this appendix, we provide additional details on how the parameters and parameter ranges used in the main text are estimated. In particular:
\begin{itemize}
    \item the procedures for selecting appropriate values for the coupling strength $\gamma$ differ between the two coupled systems considered. In the AMOC-Amazon model, the coupling strength is estimated directly from model outputs by relating changes in precipitation to changes in AMOC strength. In contrast, for the GIS-AMOC model, the coupling is directly derived from the physical structure of the model, and the way in which variation in the GIS ice volume fraction translate into freshwater forcing of the AMOC.
    \item to determine a relevant upper bound for the timescale separation~$\epsilon$, we interpret variations in~$\epsilon$ as variations in the tipping duration of the slow subsystem. Since such tipping durations are formally infinite, we adopt a finite-time approximation in which the subsystem is considered to be tipping whenever it is substantially out of equilibrium, as measured by the magnitude of its vector field. More precisely, we define the tipping duration as the time interval during which this magnitude exceeds $30\%$ of its maximum value along the transition. This definition depends on the rate of change of the external forcing applied to the slow subsystem. Accordingly, tipping durations are evaluated using the same forcing rate as in the corresponding numerical experiments.
\end{itemize}
Throughout, these parameter values and ranges should be understood as order-of-magnitude, illustrative choices intended to demonstrate how our results apply within physically reasonable regimes rather than to provide accurate estimations and bounds.

\subsection{AMOC-Amazon system}

In the Amazon model, all parameters are taken from \cite{van_nes_tipping_2014}, except for $P_d$ and $b_f$, which represent the precipitation in the absence of vegetation and the strength of the vegetation-precipitation feedback, respectively. To estimate these parameters, we partition present-day precipitation into an externally sourced component and a component recycled through vegetation. Previous studies suggest that a substantial fraction of Amazon rainfall originates from evapotranspiration within the basin, with tree transpiration accounting for about 22\% of total precipitation \cite{staal_forest-rainfall_2018}. We therefore assume that this fraction of present-day precipitation is attributable to vegetation-precipitation feedbacks. Taking mean precipitation over the Amazon to be $5\,\mathrm{mm\,day^{-1}}$, this corresponds to a vegetation-driven contribution of approximately $1.1\,\mathrm{mm\,day^{-1}}$. In the reference forest state of the model, the tree-cover fraction is $T \approx 75\%$, such that we set $b_f = 1.3\,\mathrm{mm\,day^{-1}}$, yielding $b_f (T/K) \approx 1.1\,\mathrm{mm\,day^{-1}}$ in equation~\eqref{eq:Pamaz}. The remaining externally sourced precipitation is then $P_d = 3.9\,\mathrm{mm\,day^{-1}}$.

Climate model simulations of strong AMOC weakening or collapse indicate a highly heterogeneous precipitation response over the Amazon basin, with pronounced drying in some regions and little to no change in others \cite{orihuela-pinto_interbasin_2022,nian_potential_2023,jackson_global_2015}. In particular, regional precipitation anomalies of up to about $3\ \mathrm{mm\,day^{-1}}$ are found over parts of the Amazon \cite{jackson_global_2015}. Since we only consider positive coupling values, we therefore take $0\ \mathrm{mm\,day^{-1}}$ and $3\ \mathrm{mm\,day^{-1}}$ as the lower and upper bounds of the relevant range, respectively. Noting that an AMOC collapse in our model corresponds to a change $\Delta Q \approx -3.4$, this yields a coupling strength $\gamma$ in the range $[0,\,0.88]$.

To evaluate a relevant upper limit for the timescale separation~$\epsilon$, we apply the finite-time tipping criterion defined above. For the reference value $\epsilon = 1/180$, the AMOC tipping duration is approximately $400$ years. Under strong forcing scenarios, however, model studies also show that an AMOC collapse or substantial weakening can unfold on multi-decadal timescales \cite[e.g.]{jackson_timescales_2018}. To account for this, we impose a lower bound of $50$ years on the AMOC tipping duration, which corresponds to $\epsilon \approx 0.045$. We therefore take this value as the upper limit for~$\epsilon$ in our analysis.

\subsection{GIS-AMOC system}

The coupling strength in the GIS-AMOC system is estimated by quantifying how variations in GIS ice volume influence the salinity difference between the two AMOC boxes, followed by the nondimensionalisation introduced in \cite{cessi_simple_1994}. Starting from the dimensional evolution equation for the salinity difference $\Delta S$ (eq.~(2.3) in \cite{cessi_simple_1994}), the additional meltwater input from the GIS enters as
\begin{equation}
    \frac{d}{dt}\Delta S
    = \frac{F}{H} S_0
      - \frac{\rho_\text{ice}}{\rho_\text{water}}
        \frac{2 V_\text{GIS}}{V_\text{NH}}
        \frac{dV}{dt} S_0
      - Q(\Delta \rho)\, \Delta S,
\end{equation}
where $\rho_{\text{ice}}/\rho_{\text{water}}$ converts ice loss into an equivalent freshwater flux, $V_\text{GIS}$ is the total GIS ice volume \cite{sinet_amoc_2024}, and $V_\text{NH}$ is the North Atlantic reference volume from \cite{cessi_simple_1994}. Other parameters are described and estimated in \cite{cessi_simple_1994}.

Applying the nondimensionalisation used in \cite{cessi_simple_1994} yields the expression for the coupling strength in equation~\eqref{eq:CessiVegy}, namely
\begin{equation}
    \gamma
    = 2 S_0
      \frac{\rho_\text{ice}}{\rho_\text{water}}
      \frac{\alpha_S}{\alpha_T}
      \frac{V_\text{GIS}}{\theta\, V_\text{NH}}
    \approx 3.8,
\end{equation}
which is the value adopted in our analysis.

To evaluate a relevant range for the timescale separation~$\epsilon$, we again apply the finite-time criterion. For the reference value of $\epsilon$, the GIS tipping duration is approximately $8550$ years. While GIS tipping is generally slow, model studies also show that a collapse may occur on millennial timescales under strong forcing scenarios such as RCP8.5 \cite{aschwanden_contribution_2019}. Taking $1000$ years as a conservative lower bound, this corresponds to an upper bound $\epsilon \approx 3.3$, which defines the range used in our analysis.

\section{Supplementary tables}\label{app:param}
\begin{table}[!h]
\caption{Parameters used in conceptual models. Those of the AMOC, Amazon rainforest (AR) and GIS have been taken from \cite{dijkstra_nonlinear_2013}, \cite{van_nes_tipping_2014}, and \cite{martinez_montero_surfer_2022}, respectively. In the AR model, the parameters $P_d$ and $b_f$ are estimated as explained in Appendix \ref{app:estimation}, and the dimensions of some parameters have changed with respect to \cite{van_nes_tipping_2014}, reflecting the passage to nondimensional time.} 
\label{tab:par}
\small
\vspace{6pt}
\begin{tabular}{lllll}
\hline
Model & Symbol &Name &Value& Units \\
\hline
AMOC&$\alpha$ &relaxation &$3.6\times 10^3$ & n.a.\\
&$\mu^2$ &ratio of diffusive and advective timescale &6.2 & n.a.\\
&$F_0$ &reference freshwater flux &1.1 & n.a.\\
&$\alpha_S$ &haline contraction coefficient &$0.75\times10^{-3}$ & psu$^{-1}$\\
&$\alpha_T$ &thermal expansion coefficient &$0.17\times10^{-3}$ & $^\circ$C$^{-1}$\\
&$S_0$ &reference salinity &35 & psu\\
&$\theta$ &relaxation temperature &20 & $^\circ$C\\
&$\theta_\text{diff}$ &diffusion timescale & 180& yr\\\hline
AR &$r_P$ &relaxation rate for precipitation &1.0 & n.a.\\
&$P_d$ &precipitation without vegetation &3.9 & mm day$^{-1}$\\
&$b_f$ &vegetation-precipitation feedback strength &1.32 &mm day$^{-1}$\\
&$K$ &carrying capacity &90 & \%\\
&$h_P$ &prec. where expansion rate is half of its maximum
 &0.5 & mm day$^{-1}$\\
&$r_m$ &maximum growth rate &0.3&n.a.\\
&$m_A$ &maximum mortality rate from Allee effect &0.15 &n.a.\\
&$h_A$ &threshold of tree cover for mortality incr. (Allee) &10 &\%\\
&$m_f$ &maximum mortality rate due to fire &0.11 &n.a.\\
&$h_f$ &threshold of tree cover for mortality incr. (fire) &64 &\%\\
&$\beta$ &exponent in fire term &7 & n.a.\\
&$\theta_\text{veg}$ &timescale of the Amazon model&1 & yr\\\hline
GIS &$T_+$ &warming w.r.t PI for GIS tipping point (melting)& 1.52& $^\circ$C\\
&$T_-$ &warming w.r.t PI for GIS tipping point (freezing)& 0.3 & $^\circ$C\\
&$V_+$ &GIS fract. w.r.t PI for GIS tipping point (melting)&0.77 & n.a.\\
&$V_-$ &GIS fract. w.r.t PI for GIS tipping point (freezing)&0.35 & n.a.\\
&$\delta T_0$ &reference warming  w.r.t PI&1.1 & $^\circ$C\\
&$\theta_\text{melt}$ &melting timescale &470 & yr
\end{tabular}
\vspace*{-4pt}
\end{table}

\begin{table}[!h]
\caption{Initial state variables used in conceptual models.}
\label{tab:init}
\small
\vspace{6pt}
\begin{tabular}{lllll}
\hline
Model & Symbol &Name &Value& Units \\
\hline
AMOC&$x$ &meridional temperature difference &$1.0$ & n.a.\\
&$y$ &meridional salinity difference  &0.24 & n.a.\\\hline
AR &$P$ &precipitation rate &5.0 & mm~day$^{-1}$\\
&$T$ &tree-cover fraction &75.6 & \%\\\hline
GIS &$V$ &GIS ice-volume fraction w.r.t PI& 0.90& n.a.
\end{tabular}
\vspace*{-4pt}
\end{table}
\newpage

\renewcommand\thefigure{\thesection.\arabic{figure}}    
\setcounter{figure}{0} 
\section{Supplementary figures}\label{app:supfig}
\begin{figure}[h]
    \centering
    \includegraphics[scale=1.2]{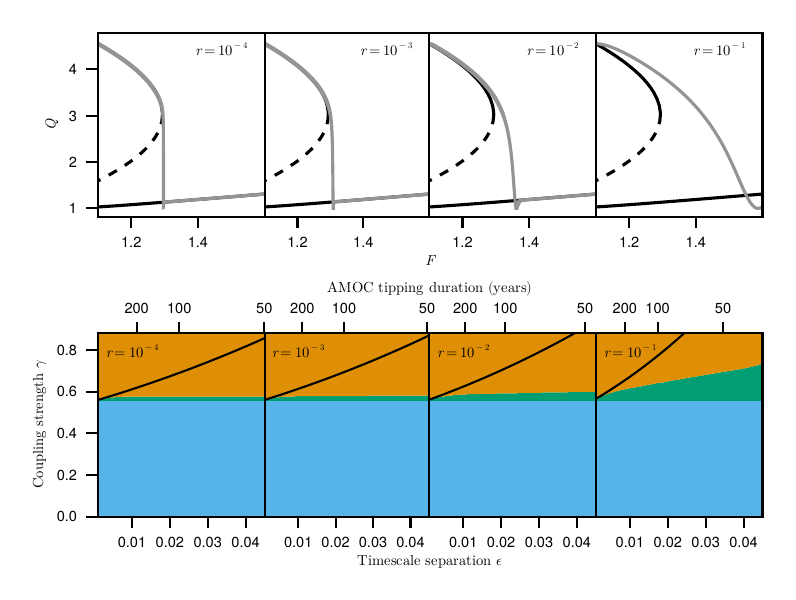}
    \caption{\emph{Effect of varying the forcing rate parameter $r$ in \eqref{eq:Fvar} on overshoot regimes of the coupled AMOC-Amazon system. The top row shows trajectories of AMOC strength $Q$ for different values of $r$, shown on the bifurcation diagram of the AMOC model, with stable (solid black) and unstable (dashed black) equilibria. The bottom row shows overshoot regimes of the AMOC-Amazon system using the same values of $r$. No-overshoot (blue), safe-overshoot (green), and unsafe-overshoot (orange) regions are shown as a function of timescale separation $\epsilon$ and coupling strength $\gamma$. The solid black lines show the approximation of the boundary between safe and unsafe overshoot provided by Result~\ref{res:1}. Because the tipping duration depends on the forcing rate, the correspondence between $\epsilon$ and AMOC tipping duration differs between panels.}
}
    \label{fig:FigS1}
\end{figure}
\begin{figure}[]
    \centering
    \includegraphics[scale=1.2]{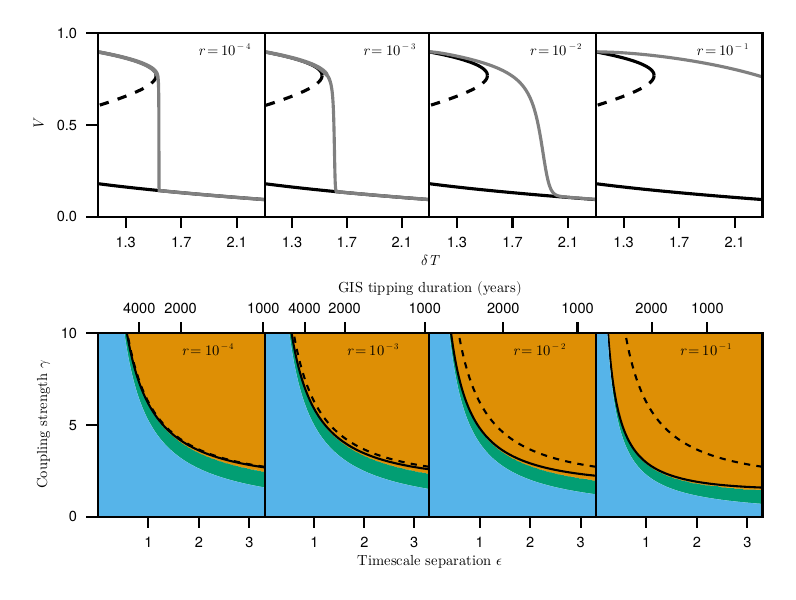}
    \caption{\emph{Effect of varying the forcing rate parameter $r$ in \eqref{eq:vardT} on overshoot regimes of the coupled GIS-AMOC system. The top row shows trajectories of GIS ice-volume fraction $V$ for different values of $r$, shown on the bifurcation diagram of the GIS model, with stable (solid black) and unstable (dashed black) equilibria. The bottom row shows overshoot regimes of the GIS-AMOC system using the same values of $r$. No-overshoot (blue), safe-overshoot (green), and unsafe-overshoot (orange) regions are shown as a function of timescale separation $\epsilon$ and coupling strength $\gamma$. The solid and dashed black lines show approximations of the boundary between safe and unsafe overshoot provided by Result~\ref{res:1} and Corollary \ref{corr:1}, respectively. Because the tipping duration depends on the forcing rate, the correspondence between $\epsilon$ and GIS tipping duration differs between panels.}}
    \label{fig:FigS2}
\end{figure}

\vskip2pc
~\newpage

\bibliographystyle{RS}
\bibliography{CascAnalytics}

~\\

\end{document}